\documentclass[% 
reprint,
superscriptaddress,
amsmath,
amssymb,aps,
prc
]{revtex4-1}

\usepackage{graphicx}% Include figure files
\usepackage{dcolumn}% Align table columns on decimal point
\usepackage{bm}% bold math
\begin{document}

\preprint{APS/123-QED}

\title{$\gamma$-ray Strength Function for Barium Isotopes}% Force line breaks with \\

\author{H. Utsunomiya}
\affiliation{Konan University, Department of Physics, 8-9-1 Okamoto, Higashinada, Japan}
\email{hiro@konan-u.ac.jp}

\author{T.~Renstr{\o}m}
\affiliation{Department of Physics, University of Oslo, N-0316 Oslo, Norway}

\author{G.~M.~Tveten}
\affiliation{Department of Physics, University of Oslo, N-0316 Oslo, Norway}

\author{S.~Goriely}
\affiliation{Institut d'Astronomie et d'Astrophysique, Universit\'{e} Libre de Bruxelles, Campus de la Plaine, CP-226, 1050 Brussels, Belgium}

\author{T.~Ari-izumi}
\affiliation{Konan University, Department of Physics, 8-9-1 Okamoto, Higashinada, Japan}

\author{V.W. Ingeberg}
\affiliation{Department of Physics, University of Oslo, N-0316 Oslo, Norway}

\author{B. V.~Kheswa}
\affiliation{Department of Physics, University of Oslo, N-0316 Oslo, Norway}
\affiliation{University of Johannesburg, Department of Applied physics and Engineering mathematics, Doornfontein, Johannesburg, 2028, South Africa}

\author{Y.-W.~Lui}
\affiliation{Cyclotron Institute, Texas A\& M University, College Station, Texas 77843, USA}

\author{S.~Miyamoto}
\affiliation{Laboratory of Advanced Science and Technology for Industry, University of Hyogo, 3-1-2 Kouto, Kamigori, Ako-gun, Hyogo 678-1205, Japan}

\author{S.~Hilaire}
\affiliation{CEA, DAM, DIF, F-91297 Arpajon, France}

\author{S.~P\'{e}ru}
\affiliation{CEA, DAM, DIF, F-91297 Arpajon, France}

\author{A.~J.~Koning}
\affiliation{Nuclear Data Section, International Atomic Energy Agency, A-1400 Vienna, Austria}

\date{\today}% It is always \today, today,

\begin{abstract}
Photoneutron cross sections were measured for $^{137}$Ba and $^{138}$Ba at energies below two-neutron threshold using quasi-monochromatic $\gamma$-ray beams produced in laser Compton-scattering at the NewSUBARU synchrotron radiation facility. 
The photoneutron data are used to constrain the $\gamma$-ray strength function on the basis of the Hartree-Fock-Bogolyubov plus quasi-particle
random phase approximation using the Gogny D1M interaction.
Supplementing the experimentally constrained $\gamma$-ray strength function with the zero-limit E1 and M1 contributions which are unique to the deexcitation mode, we discuss radiative neutron capture cross sections relevant to the s-process nucleosynthesis of barium isotopes in the vicinity of the neutron magic number 82.  

\end{abstract}

%\pacs{Valid PACS appear here}
\maketitle
\section{Introduction}
The $\gamma$-ray strength function ($\gamma$SF) \cite{Bartholomew73,Lone85} is a statistical quantity employed in the Hauser-Feshbach model of the compound nuclear reaction. The $\gamma$SF in the de-excitation mode is a key quantity to determine radiative neutron capture cross sections that are of direct relevance to the s-process nucleosynthesis of elements heavier than iron. The downward $\gamma$SF for dipole radiation with a given energy $\varepsilon_{\gamma}$ is defined \cite{RIPL3} by 

\begin{equation}
\overleftarrow{f_{X1}}(\varepsilon_{\gamma})=\varepsilon_{\gamma}^{-3} \frac{\langle\Gamma_{X1}(\varepsilon_{\gamma})\rangle}{D_{\ell}}.
\label{eq:dgsf}
\end{equation}
 
 \noindent 
Here  $X$ is either electric ($E$) or magnetic ($M$),  $\langle\Gamma_{X1}(\varepsilon_{\gamma})\rangle$ is the average radiation width, and $D_{\ell}$ is the average level spacing for s-wave ($\ell$=0) or p-wave ($\ell$=1) neutron resonances.  
 
In contrast, the $\gamma$SF in the excitation mode for dipole radiation is defined \cite{RIPL3} by the average  cross section for $E1/M1$ photoabsorption 
$\langle\sigma_{X1}(\varepsilon_{\gamma})\rangle$ to the final states with all possible spins and parities \cite{Lone85}:

\begin{equation}
\overrightarrow{f_{X1}}(\varepsilon_\gamma)=\frac{\varepsilon_{\gamma}^{-1}}{3(\pi\hbar c)^2} \langle\sigma_{X1}(\varepsilon_{\gamma})\rangle.
\label{eq:ugsf}
\end{equation} 
 
Above neutron separation energy except at energies near neutron threshold, the total upward $\gamma$SF can be determined by substituting 
$\langle\sigma_{X1}(\varepsilon_{\gamma})\rangle$ with experimental photoneutron cross sections that dominate photoabsorption cross sections.    

A recent systematic study across the chart of nuclei has formulated the zero-limit behavior of both E1 and M1 strengths in the analytical form \cite{Goriely18a}, the latter of  which (referred to as M1 upbend) was experimentally observed \cite{Voin04,Gutt05,Algi08} and theoretically supported by the shell-model calculation \cite{Schw13,Brow14,Siej17a,Siej17b,Kara17,Schw17}. The presence of the zero-limit strengths which correspond to $\gamma$-ray transitions between high-lying states is unique to the downward $\gamma$SF, showing that the Brink hypothesis of the approximate equality of $\overleftarrow{f_{X1}}$ and $\overrightarrow{f_{X1}}$ \cite{Brink,Axel} is violated.  

The radiative neutron capture in the s-process path which proceeds along the valley of $\beta$ stability successively produce heavier isotopes by adding one neutron at a time until it is intervened by $\beta^{-}$ decay which increases the atomic number of the element by one.  When the s-process flow reaches a magic neutron number, the nuclear binding energy increases by an amount of the order of 1 - 2 MeV, leading to a drop of the neutron separation energy of a compound nucleus which is formed by the neutron capture. The drop results in large level spacing $D$ in Eq.~(\ref{eq:dgsf}), and significantly decreases the downward $\gamma$SF and radiative neutron capture cross section of neutron-magic nuclei \cite{BW79}.  As a result, the s-process nucleosynthesis forms the 1st, 2nd, and 3rd peaks of elements around the mass number $A$ = 86--90, 138--142 and 208, corresponding to the magic number 50, 82, and 126, respectively.
 
In this paper, we discuss radiative neutron capture cross sections for barium isotopes including $^{138}$Ba with the neutron magic number 82 in terms of the deexcitation $\gamma$SF based on the $\gamma$-ray strength function method \cite{Utsu18a, Utsu19}. 

The $\gamma$SF from two relatively different nuclear models are studied, namely the Simple Modified Lorentzian (SMLO) model \cite{Goriely18b} and the Hartree-Fock-Bogolyubov plus quasi-particle random phase approximation (QRPA) based on the Gogny D1M interaction (hereafter denoted as D1M+QRPA) \cite{Martini16,Goriely16b,Goriely18a} for both E1 and M1 components is constrained to the present photoneutron cross section measured for $^{138}$Ba and, for the first time, $^{137}$Ba. The experimentally constrained $\gamma$SF is further supplemented with the zero-limit E1 and M1 contributions which are unique to the deexcitation mode. 

\begin{figure*}
\includegraphics[bb = 150 100 530 425, scale=0.55]{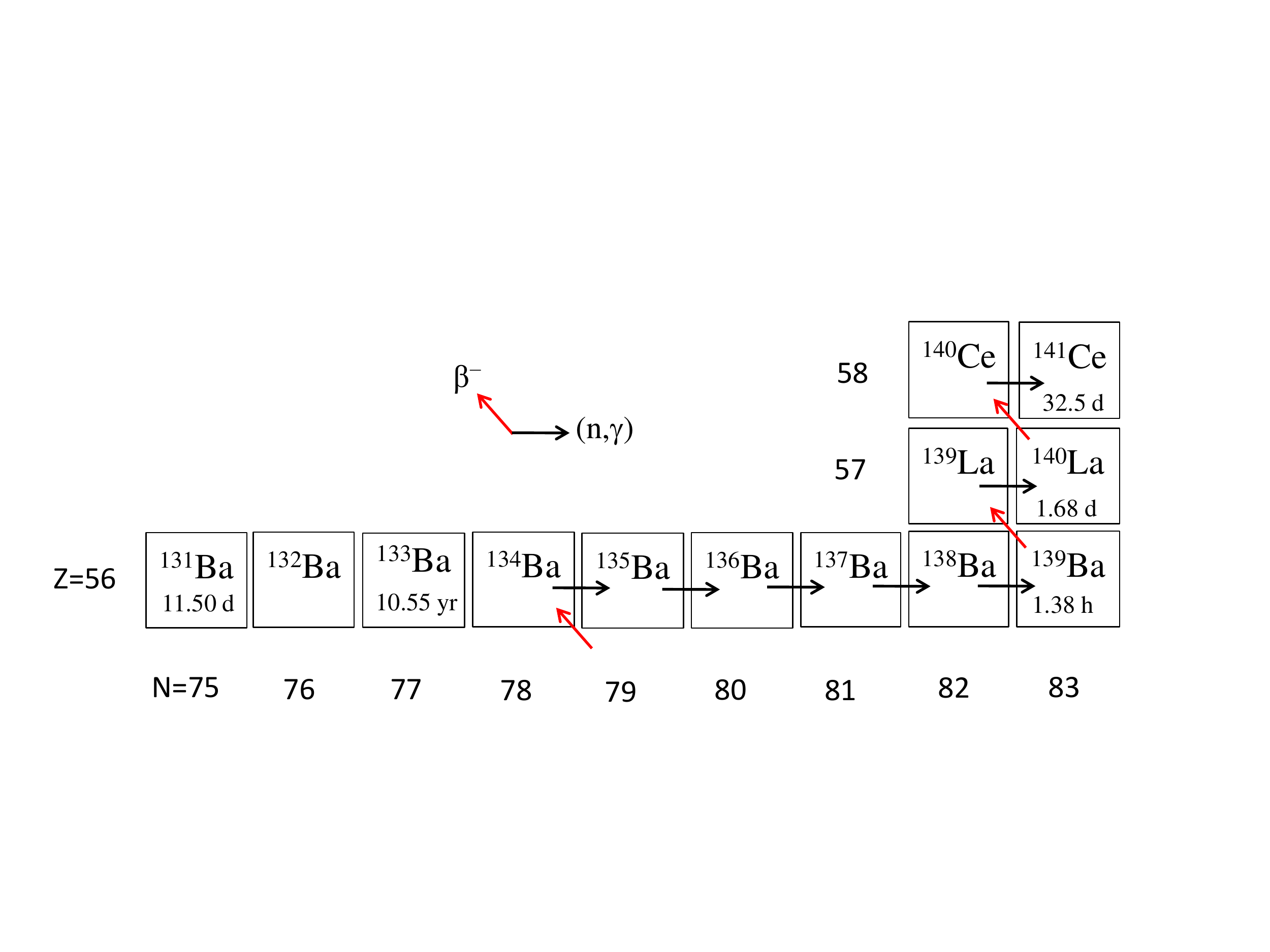}
\caption{(Color online)  An excerpt of the chart of nuclei depicting the Ba region along the s-process path.}
\label{fig:chart}
\end{figure*}

Based on the same nuclear ingredients, radiative neutron capture cross sections are calculated with the TALYS code \cite{TALYS} for other stable barium isotopes including $^{138}$Ba and compared with experimental data.
The Ba isotopes of the present research interest along the s-process path are shown in Fig.~\ref{fig:chart}.

In Sect.~\ref{sec_exp}, our experimental procedure is described and in Sect.~\ref{sec_data} data are analyzed. Our resulting photoneutron cross sections are discussed in Sect.~\ref{sec_disc} and compared with D1M+QRPA calculations.  Radiative neutron capture cross sections calculated for stable barium isotopes are compared with experimental data. Finally conclusions are drawn in Sect.~\ref{sec_conc}.

\section{Experimental procedure} 
\label{sec_exp}
The photo-neutron measurements on $^{137,138}$Ba took place at the NewSUBARU synchrotronic radiation facility. 
Quasi-monochromatic pencil-like $\gamma$-ray beams were produced through laser Compton scattering (LCS) of 1064 nm photons from the Nd:YVO$_4$ laser in head-on collisions with relativistic electrons. The electrons were injected from a linear accelerator into the NewSUBARU storage ring with an initial energy of 974 MeV, then subsequently decelerated to nominal energies in the region from 620 and 694~MeV to 849~MeV, providing LCS $\gamma$-ray beams corresponding to the neutron separation energy $S_n$ up to 13 MeV for $^{137}$Ba and $^{138}$Ba respectively. In total, 22 and 15 individual $\gamma$ beams were produced for both $^{137}$Ba and $^{138}$Ba, respectively. 

The $^{137,138}$Ba targets were made form isotopically enriched barium carbonate (BaCO$_3$). The material was in pressed together and enclosed in an Al-cylinder with a thin cap. The targets had  areal density of $2.236$~g/cm$^2$ and $5.642$~g/cm$^2$, for $^{137}$Ba and $^{138}$Ba, respectively. The presence of the Al-cap made it necessary not to increase the $\gamma$-ray beam energy above 13 MeV due to $S_n$=13.056 MeV for $^{27}$Al. The corresponding enrichment of the two isotopes were $85.0\%$ and $99.9\%$.  The main contaminant of the $^{137}$Ba-target was $^{138}$Ba ($\sim$~15$\%$). This was taken into account by using only the number of  $^{137}$Ba nuclei below $S_n$ of $^{138}$Ba, 8611.72 keV, and by considering the neutron contribution from $^{138}$Ba above.  

The LCS $\gamma$-ray beam produced at NewSUBARU is well calibrated in absolute energy \cite{Utsu14}, energy profile \cite{Ioana_thesis,Fili14,Utsu15}, and flux \cite{Kondo2011,Utsu18b}.  One can find more details of generation and calibration of the LCS $\gamma$-ray beam, and detection of neutrons with a high-efficiency moderation-based detector \cite{neutrondet} in Ref.~\cite{Utsu19}. 

\begin{figure}[t]
\includegraphics[width=0.5\textwidth]{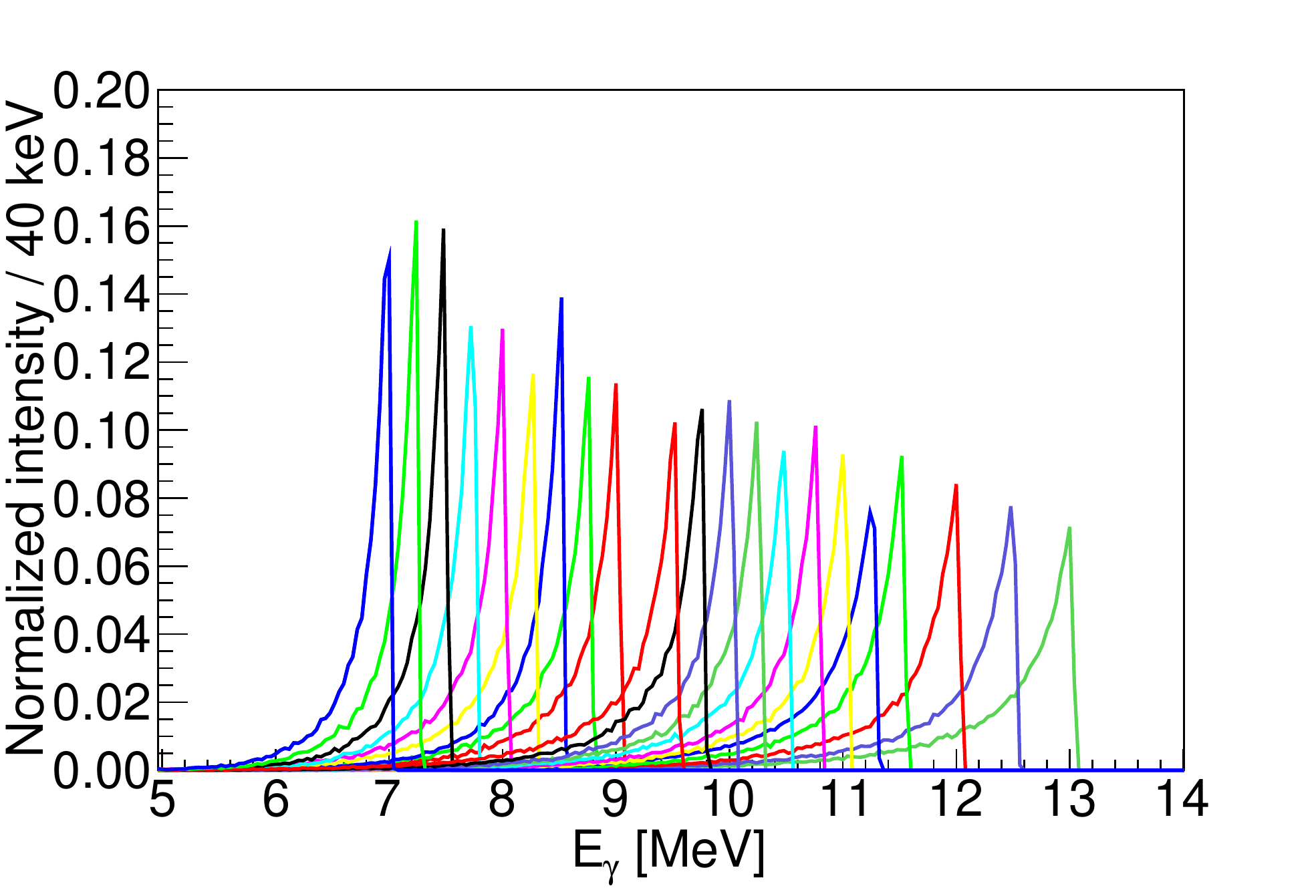}
\caption{(Color online) The simulated energy profiles for the $\gamma$-beams used in the $^{137,138}$Ba measurements. The distributions are normalized to unity. }
\label{fig:GammaProfile}
\end{figure}

The measured photo-neutron cross section for an incoming beam with maximum $\gamma$-energy $E_{\rm max}$ is given by the convoluted cross section,
\begin{equation}
\sigma^{E_{\rm max}}_{\rm exp}=\int_{S_n}^{E_{\rm max}}D^{E_{\rm max}}(E_{\gamma})\sigma(E_{\gamma})dE_{\gamma}=\frac{N_n}{N_tN_{\gamma}\xi\epsilon_n g}.
\label{eq:cross1}
\end{equation}
Here, $D^{E_{\rm max}}$ is the normalized,$\int_{S_n}^{E_{\rm max}} D^{E_{\rm max}}dE_{\gamma}= 1$, energy distribution of
the $\gamma$-ray beam obtained from a GEANT4 \cite{geant4ref} simulation analysis of experimental response functions of a LaBr$_3$(Ce) beam-profile monitor. The simulated profiles of the $\gamma$ beams, $D^{E_{\rm max}}$, used to investigate $^{137}$Ba are shown in Fig.~\ref{fig:GammaProfile}. Furthermore, $\sigma(E_{\gamma})$ is the true photo-neutron cross section as a function of energy. The quantity $N_n$ represents the number of neutrons detected, $N_t$ gives the number of target nuclei per unit area, $N_{\gamma}$ is the number of $\gamma$ rays incident on target, $\epsilon_n$ represents the neutron detection efficiency, and finally $\xi=(1-e^{-\mu t})/(\mu t)$ gives a correction factor for self-attenuation in the target. The factor $g$ represents the fraction of the $\gamma$ flux above $S_n$. 

We have determined the convoluted cross sections $\sigma^{E_{\rm max}}_{\rm exp}$ given by Eq.~(\ref{eq:cross1}) for $\gamma$ beams with maximum energies in the range $S_{n}\leq E_{\rm max} \leq$ 13 MeV. The convoluted cross sections $\sigma^{E_{\rm max}}_{\rm exp}$ are not connected to a specific $E_{\gamma}$, and we choose to plot them as a function of $E_{\gamma \rm max}$. The convoluted cross sections of the two Ba-isotopes, which are often called monochromatic cross sections, are shown in Fig.~\ref{fig:MonocrossBoth}. The error bars in Fig.~\ref{fig:MonocrossBoth} represent the total uncertainty in the quantities comprising Eq.~(\ref{eq:cross1}) and consists of $\sim 3.2\%$ from the efficiency of the neutron detector, $\sim 1\%$ from the pile-up method that gives the number of $\gamma$-rays, and the statistical uncertainty in the number of detected neutrons. The statistical error ranges between $\sim$ 13 $\%$ close to neutron threshold and 4.6 $\%$ for the highest maximum $\gamma$-ray beam energies. Except for the first few data points close to separation energy, the total error is dominated by the uncertainty stemming from the pile-up method and from the simulated efficiency of the neutron detector.
For the total uncertainty, we have added these uncorrelated errors quadratically.

%---------------------------------------------------%
%---------------------------------------------------%

\section{Data analysis}
\label{sec_data}
Now we extract the deconvoluted, $E_{\gamma}$ dependent, photo-neutron cross section, $\sigma(E_{\gamma})$, from the integral of Eq.~(\ref{eq:cross1}) \cite{Rens18}.
Each of the measurements characterized by the beam energy, $E_{\rm max}$, corresponds to folding of $\sigma(E_{\gamma})$ with the 
measured beam profile, $D^{E_{\rm max}}$.  
%---------------------------------------------------%
\begin{figure}[]

\includegraphics[width=0.48\textwidth]{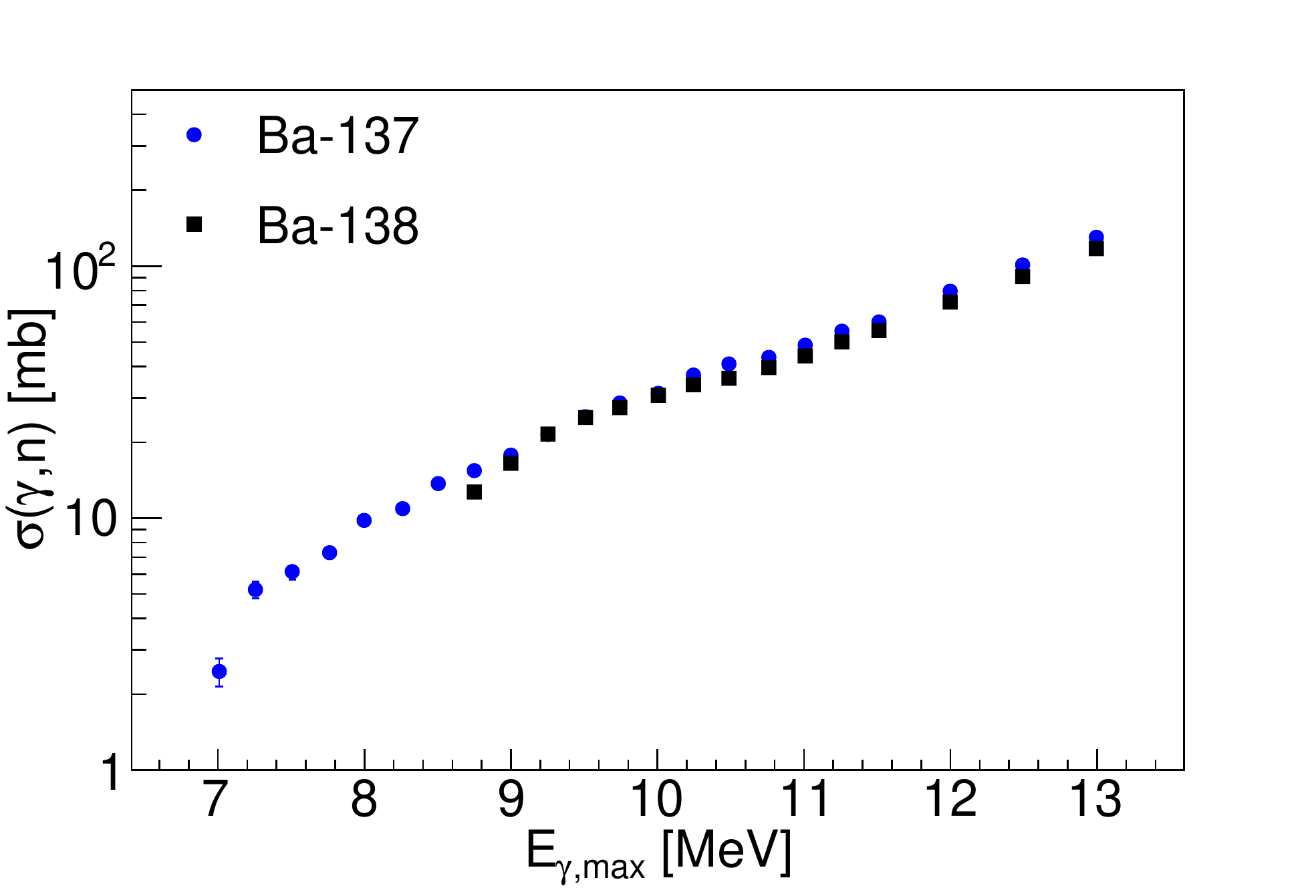}
\caption{(Color online) Monochromatic cross sections of $^{137}\rm{Ba}$ and $^{138}\rm{Ba}$. The error bars contain statistical uncertainties from the number of detected neutrons, the uncertainty in the efficiency of the neutron detector and the uncertainly in the pile-up method used to determine the number of incoming $\gamma$'s on target.}
\label{fig:MonocrossBoth}
\end{figure}
%---------------------------------------------------%
By approximating the integral in Eq.~(\ref{eq:cross1}) with a sum for each $\gamma$-beam profile, we are able to express the problem as a set of linear equations
\begin{equation}
\sigma_{\rm f }=\bf{D}\sigma,
\end{equation}
where $\sigma_{\rm f}$ is the cross section folded with the beam profile {\bf D}.  
The indexes $i$ and $j$ of the matrix element $D_{i,j}$ corresponds to $E_{\rm max}$ and $E_{\gamma}$, respectively.
The set of equations is given by
\begin{equation}
\begin{pmatrix}\sigma_{\rm{1}}\\\sigma_{\rm{2}}\\ \vdots \\ \sigma_N \end{pmatrix}_{\rm f}\\\mbox{}=
\begin{pmatrix}D_{ 11} & D_{ 12} & \cdots &\cdots &D_{ 1M} \\ D_{ 21} & D_{ 22} &
\cdots & \cdots & D_{ 2M} \\ \vdots & \vdots & \vdots & \vdots & \vdots \\ D_{ N1} & D_{ N2}& \cdots & \cdots &D_{ NM}\end{pmatrix}
\begin{pmatrix}\sigma_{1}\\\sigma_{2}\\ \vdots \\ \vdots \\\sigma_{M} \end{pmatrix}.
\label{eq:matrise_unfolding}
\end{equation}
Each row of $\bf{D}$ corresponds to a GEANT4 simulated $\gamma$
beam profile belonging to a specific measurement characterized by $E_{\rm max}$.  See Fig.~\ref{fig:GammaProfile} for a visual representation of the response matrix $\bf{D}$. It is clear that $\bf{D}$ is highly asymmetrical.

As mentioned, we have used $N=15$ beam energies when investigating $^{137}$Ba and 18 for $^{138}$Ba, but the beam profiles above $S_n$ is simulated in steps of 10 keV in $\gamma$-ray energy. As the system of linear equations in Eq.~(\ref{eq:matrise_unfolding}) is under-determined, the true $\sigma$ vector cannot be extracted by matrix inversion. In order to find $\sigma$, we utilize a folding iteration method. The main features of this method are as follows:

\begin{itemize}

\item [1)] As a starting point, we choose for the 0th iteration, a constant trial function $\sigma^0$.
This initial vector is multiplied with $\bf{D}$, and we get the 0th folded vector $\sigma^0_{\rm f}= {\bf D} \sigma^{0}$.
\item[2)] The next trial input function, $\sigma^1$, can be established by adding the difference of the experimentally measured spectrum, $\sigma_{\rm{exp}}$, and the folded spectrum, $\sigma^0 _{\rm f}$,
to $\sigma^0$. In order to be able to add the folded and the input vector together, we first perform a  Piecewise Cubic Hermite Interpolating Polynomial (pchip) interpolation on the folded vector so that the two vectors have equal dimensions. Our new input vector is:

\begin{equation}
\sigma^1 = \sigma^0 + (\sigma_{\rm{exp}}-\sigma^0 _{\rm f}).
\end{equation} 

\item[3)] The steps 1) and 2) are iterated $i$ times giving
\begin{eqnarray}
\sigma^i_{\rm f} &=& {\bf D} \sigma^{i}
\\
\sigma^{i+1}     &=& \sigma^i + (\sigma_{\rm{exp}}-\sigma^i _{\rm f})
\end{eqnarray}
until convergence is achieved. This means that
$\sigma^{i+1}_{\rm f} \approx \sigma_{\rm exp}$ within the statistical errors.
In order to quantitatively check convergence, we calculate the reduced $\chi^2$ of $\sigma^{i+1}_{\rm f}$ and
$\sigma_{\rm{exp}}$ after each iteration.
Approximately four iterations are usually enough for convergence, which is defined when the reduced $\chi^2$ value approaches $\approx 1$.
\end{itemize}

We stopped iterating when the $\chi^2$ started to
be lower than unity. In principle, the iteration could continue until the reduced $\chi^2$ approaches zero,
but that results in large unrealistic fluctuations in $\sigma^i$ due to over-fitting to the measured points $\sigma_{\rm exp}$.

\begin{figure}[]
\includegraphics[width=0.48\textwidth]{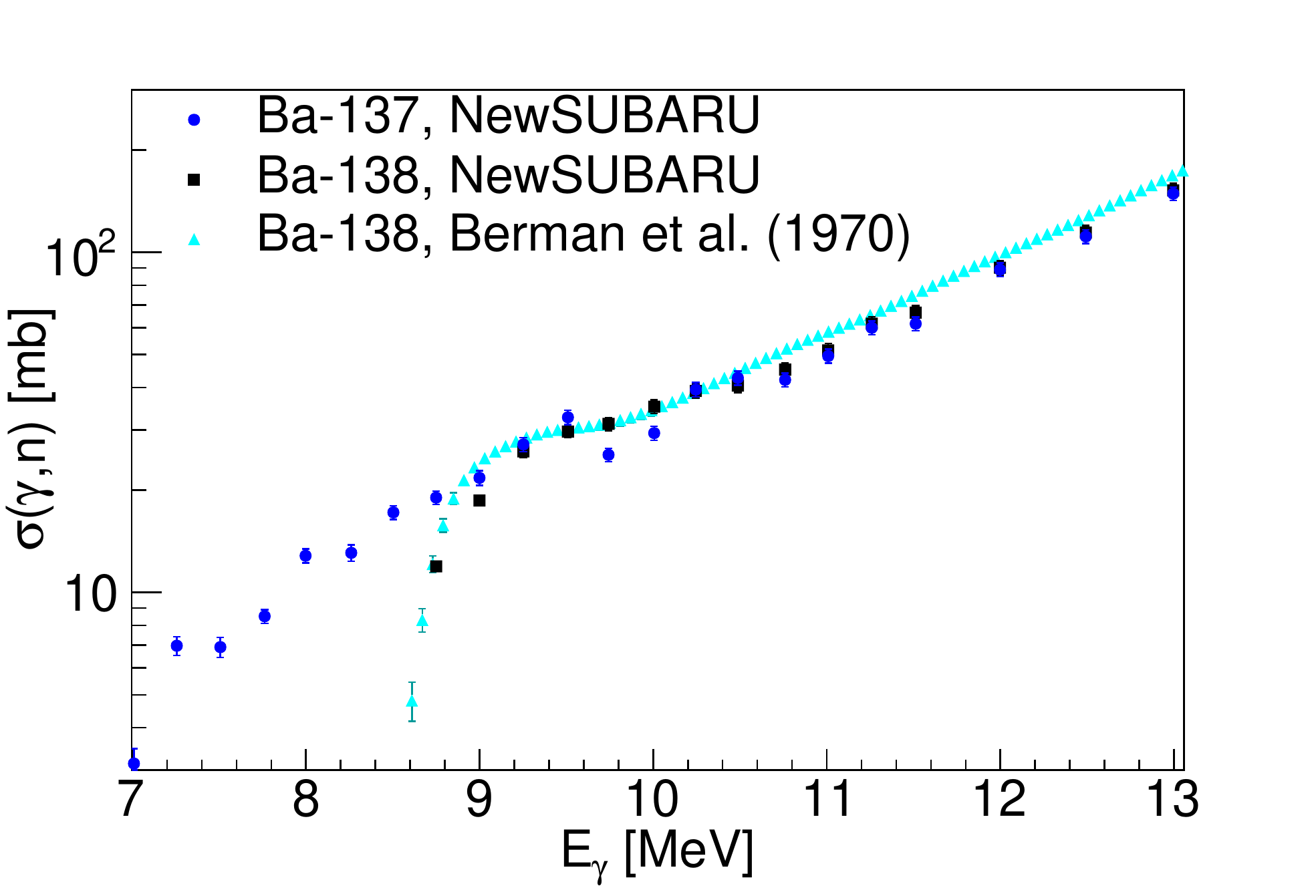}
\caption{(Color online) Unfolded cross sections of $^{137,138}\rm{Ba}$ in comparison with the previous measurements  for $^{138}\rm{Ba}$ \cite{Berman70}.}
\label{fig:Unfolded}
\end{figure}

We estimate the total uncertainty in the unfolded cross sections by calculating an upper limit of the monochromatic cross sections from Fig.\ref{fig:MonocrossBoth} by adding and subtracting the errors to the measured cross section values. This upper and lower limit is then unfolded separately, resulting in the unfolded cross sections shown in Fig.~\ref{fig:Unfolded}.

In Fig.~\ref{fig:Unfolded}, the unfolded cross sections for $^{137}$Ba and $^{138}$Ba are evaluated at the maximum energies of the incoming $\gamma$ beams. The error bars represent the difference between the upper and lower limit of the unfolded cross sections.  
%The unfolded cross sections are tabulated in Table I.

%---------------------------------------------------%
%---------------------------------------------------%

\begin{table}[tb]
\centering
\caption{The unfolded cross sections $\sigma$ evaluated at the maximum $\gamma$-ray energy and the total (statistical plus systematic) uncertainty $\Delta \sigma$. }
\label{tab:unfoldedtab}
\begin{tabular}{llll}
 \hline
$^{137}$Ba \hspace{5mm}  & $E_{\gamma}$ [MeV]& $\sigma$ [mb] & $\Delta \sigma$ [mb] \\
 \hline
 & 7.01 & 3.14	  &  0.33  \\
 & 7.26 & 6.97	  &  0.44  \\
 & 7.51 & 6.90	  &  0.47  \\
 & 7.76 & 8.51	  &  0.40  \\
 & 8.00 & 12.81	  &  0.62  \\
 & 8.26 & 13.07	  &  0.72  \\
 & 8.50 & 17.18  &  0.80  \\
 & 8.75 & 18.99 &  0.87  \\
 & 9.00 & 21,71 &  1.08  \\
 & 9.25 & 27.22 &  1.30  \\
 & 9.51 & 32.71 &  1.60  \\
 & 9.74 & 25.38 &  1.16  \\
 & 10.01 & 29.40 &  1.40 \\
 & 10.25 & 39.46 &  1.89 \\
 & 10.49 & 42.70 &  2.06 \\ 
 & 10.76 & 42.14 &  1.95 \\
 & 11.01 & 49.51 &  2.37 \\
 & 11.26 & 59.97 &  2.79 \\
 & 11.51 & 61.64 &  2.88 \\
 & 12.00 & 88.85 &  4.11 \\ 
 & 12.49 & 111.26 &  5.10 \\
 & 13.00 & 148.72 &  6.84 \\
  \hline
 $^{138}$Ba \hspace{5mm} & $E_{\gamma}$ [MeV]& $\sigma$ [mb] & $\Delta \sigma$ [mb] \\
   \hline
 & 8.75 & 11.93	  &  0.27 \\
 & 9.00 & 18.62	  &  0.62 \\
 & 9.25 & 25.97	  &  1.00 \\
 & 9.51 & 29.72	  &  1.21 \\
 & 9.74 & 31.28	  &  1.39 \\
 & 10.01 & 35.16 &  1.63 \\
 & 10.25 & 39.08 &  1.82 \\
 & 10.49 & 40.57  &  1.90 \\
 & 10.76 & 45.16 &  2.15 \\
 & 11.01 & 51.48 &  2.47 \\
 & 11.26 & 61.69 &  2.93 \\
 & 11.51 & 66.46 &  3.20 \\
 & 12.00 & 90.04 &  4.35 \\
 & 12.49 & 114.18 &  5.52 \\
 & 13.00 & 151.95 &  7.36 \\
  \hline
\end{tabular}
\end{table}

\section{Discussion}
\label{sec_disc}

%---------------------------------------------------%
\begin{figure}
\includegraphics[width=1.0\columnwidth]{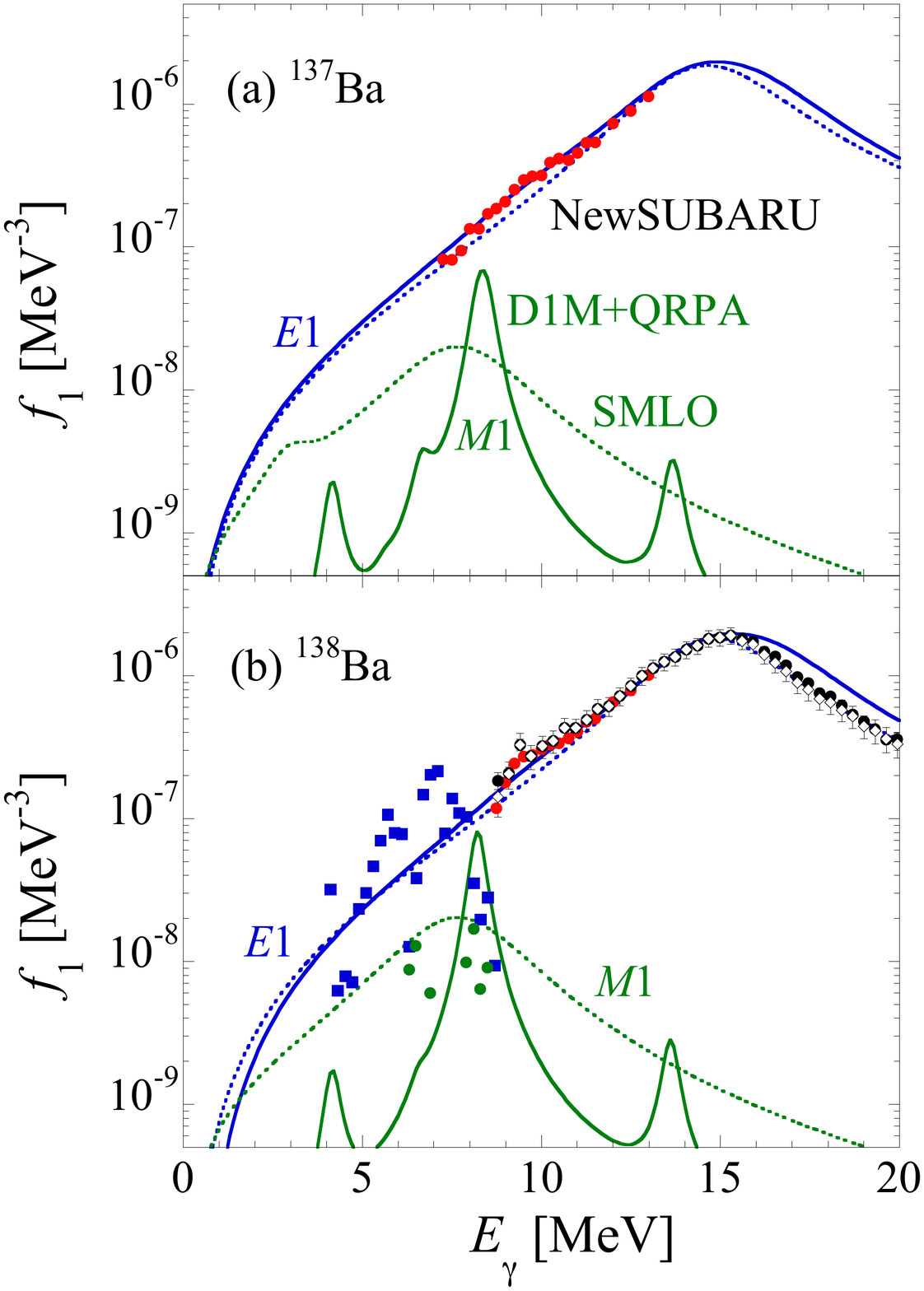}
\caption{(Color online) (a) Comparison of the D1M+QRPA (solid lines) and SMLO (dotted lines) $\gamma$SF  for $^{137}\rm{Ba}$ with the measured strength function extracted from the present NewSUBARU experiment (red circles). The $E1$ mode is shown by blue lines and $M1$ by green lines. 
(b) same for and $^{138}\rm{Ba}$ $\gamma$SF. Previous measurements from nuclear resonance fluorescence \cite{Tonchev10} for the $E1$ (blue squares) and $M1$ (green circles) modes, as well as photoneutron data (solid circles) \cite{Berman70} and its evaluation (open diamonds)  \cite{Varlamov16} are also shown.}
\label{fig:psf}
\end{figure}
%---------------------------------------------------%

%---------------------------------------------------%
\begin{figure}
\includegraphics[width=1.0\columnwidth]{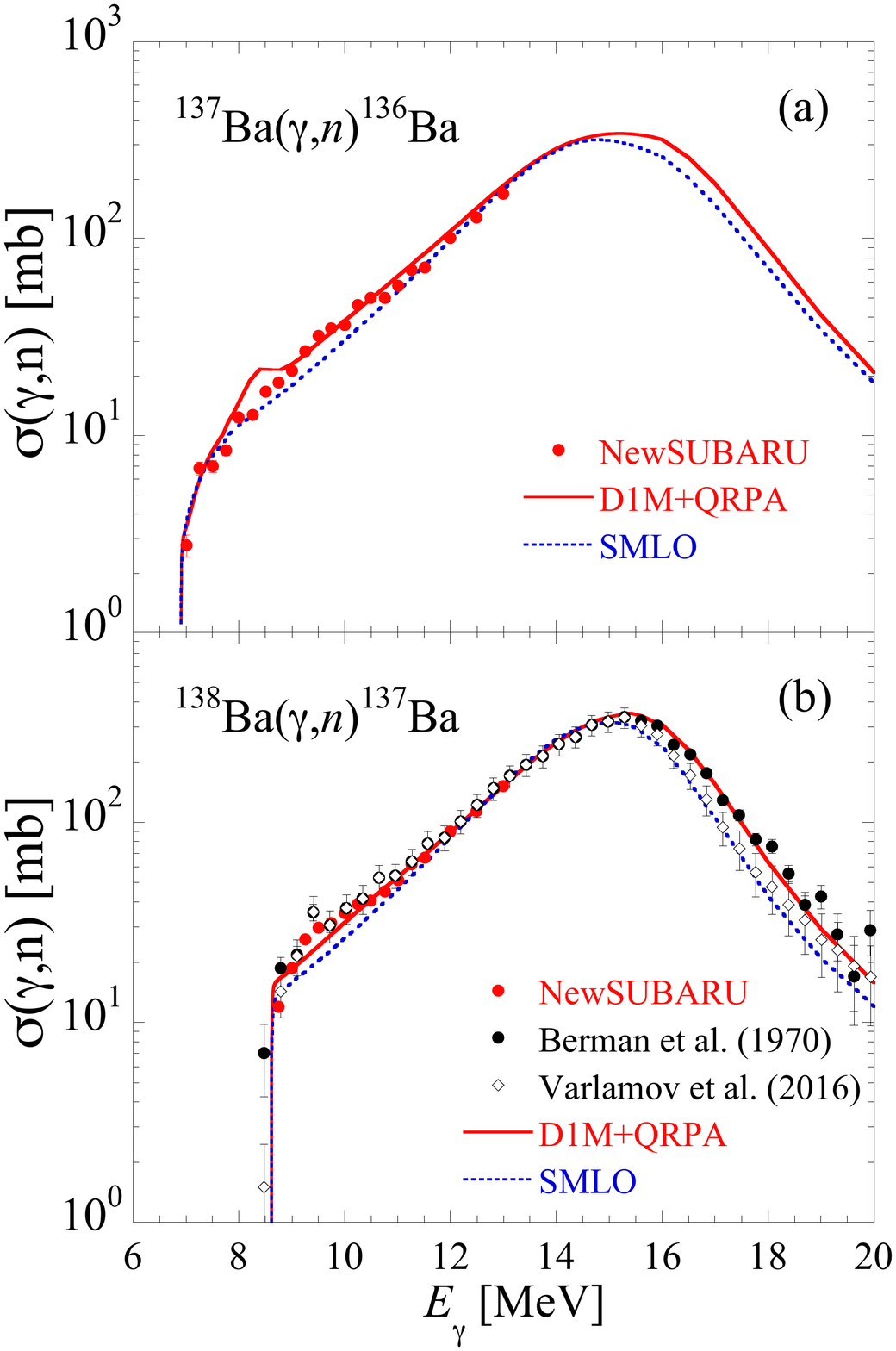}
\caption{(Color online) (a) Present $^{137}\rm{Ba}$($\gamma$,n)$^{136}\rm{Ba}$ measured cross sections compared with the TALYS calculations based on the SMLO (dashed blue line) and D1M+QRPA (solid red line) $\gamma$SF. (b) same for  $^{138}\rm{Ba}$($\gamma$,n)$^{137}\rm{Ba}$ reaction where previous measurements (black circles) \cite{Berman70} and evaluation (open diamonds)  \cite{Varlamov16} are also shown.}
\label{fig:gn}
\end{figure}
%---------------------------------------------------%
%---------------------------------------------------%
\begin{figure}
\includegraphics[width=1.0\columnwidth]{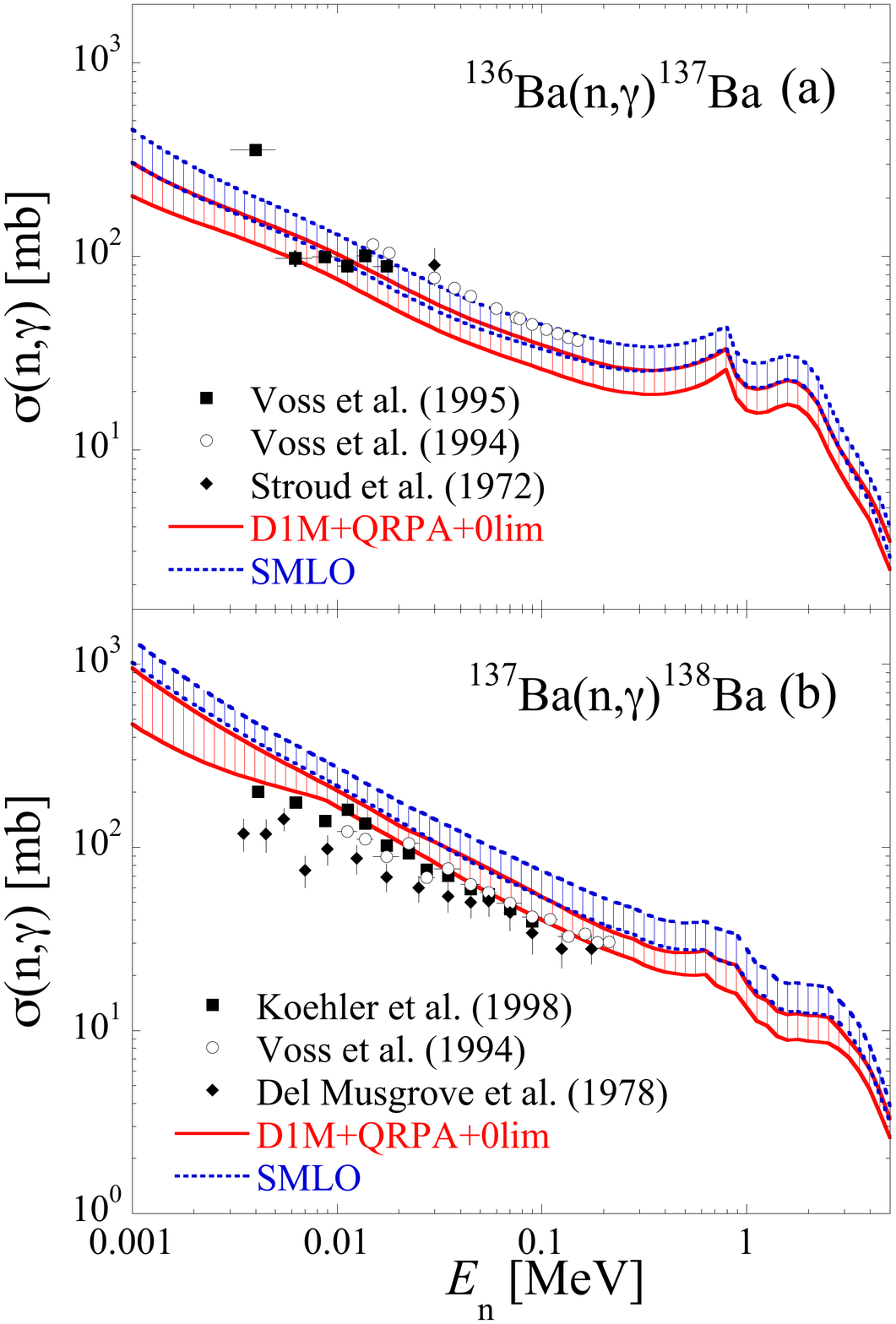}
\caption{(Color online) (a) Comparison of the $^{136}\rm{Ba}$(n,$\gamma$)$^{137}\rm{Ba}$ measured cross sections \cite{Stroud72,Voss95,Voss94} with the TALYS calculations based on the SMLO (dashed blue line) and D1M+QRPA+0lim (solid red line) $\gamma$SF. (b) same for  $^{137}\rm{Ba}$(n,$\gamma$)$^{138}\rm{Ba}$ cross section \cite{Voss94,Koehler98,Musgrove78}. Theoretical uncertainties correspond to the use of different level density models  \cite{Koning08,Goriely08}.}
\label{fig:ng}
\end{figure}
%---------------------------------------------------%

The present experimental results are now analyzed in light of the recent systematics of the $\gamma$SF obtained within the phenomenological Lorenztian approach with the SMLO model \cite{Goriely18b} and within the mean field plus QRPA calculations with the D1M+QRPA model \cite{Martini16,Goriely16b,Goriely18a}. While the SMLO E1 strength is essentially fitted on available photoabsorption data, the D1M+QRPA has only globally be renormalized to reproduce the bulk of experimental data. Such a renormalization includes a broadening of the QRPA strength  to take the neglected damping of  collective motions into account as well as a shift of the strength to lower energies due to the contribution beyond the 1 particle - 1 hole excitations and the interaction between the single-particle and low-lying collective phonon degrees of freedom. Such phenomenological corrections  have been applied to the present  Ba isotopes, as described in Ref.~\cite{Goriely18a}. In addition, in order to reproduce the present photoneutron cross section in the low-energy tail of the giant dipole resonance, we find that a global energy shift of 0.5~MeV of the overall $E1$ strength is required in the specific case of $^{138}$Ba. Such renormalizations are within the uncertainties affecting the $\gamma$SF predictions \cite{Goriely18a}. The resulting D1M+QRPA and SMLO $\gamma$SF  for $^{137}\rm{Ba}$ and $^{138}\rm{Ba}$ are shown in Fig.~\ref{fig:psf} and compare with the present and previous data. In the case of  $^{138}\rm{Ba}$, nuclear resonance fluorescence measurements \cite{Tonchev10} are also available separately for the $E1$ and $M1$ modes. Details on the extraction of the $\gamma$SF from measured data can be found in Ref.~\cite{Goriely19}.

Both the D1M+QRPA and SMLO models have proven their capacity to reproduce relatively well all types of experimental data bearing relevant information on $\gamma$SF, such as those extracted from photoabsorption, nuclear resonance fluorescence, Oslo method, radiative neutron or proton captures or inelastic proton experiments  as shown in Ref.~\cite{Goriely19}. We show in Figs.~~\ref{fig:psf}-\ref{fig:gn} that this is also the case for the present Ba photoneutron data in the whole energy range covered by our experiment. In the $^{137}\rm{Ba}$($\gamma$,n)$^{136}\rm{Ba}$ case, the D1M+QRPA M1 spin-flip strength at 8.5~MeV is however clearly overestimated, while the  SMLO model underestimates the $E1$ strength in the 9-11 MeV region. 

Another way of testing our photoneutron data and the $\gamma$SF deduced out of those is to consider the reverse radiative neutron capture cross sections. Those are available for $^{136}\rm{Ba}$ and $^{137}\rm{Ba}$ and therefore are sensitive to the low-energy $\gamma$SF of $^{137}\rm{Ba}$ and $^{138}\rm{Ba}$, respectively.  We have considered both the SMLO and D1M+QRPA $E1$ and $M1$ strengths as described above and applied them to the TALYS calculation of the (n,$\gamma$) cross section. It should however be stressed that the photoabsorption strength needs to be complemented by the zero-limit correction when considering the de-excitation of the compound nucleus formed by the neutron capture. Inspired from shell model studies, this low-energy limit has been approximated in Ref.~\cite{Goriely18a} for D1M+QRPA (the model is then referred to as D1M+QRPA+0lim) and in Ref.~\cite{Goriely18b} for SMLO. The radiative cross sections are also sensitive to the adopted nuclear level density model. For these reasons, different prescriptions \cite{Koning08,Goriely08} have been considered in the TALYS calculations.

We compare in Fig.~\ref{fig:ng}  the $^{136}\rm{Ba}$(n,$\gamma$)$^{137}\rm{Ba}$ and $^{137}\rm{Ba}$(n,$\gamma$)$^{138}\rm{Ba}$ measured cross sections with the TALYS Hauser-Feshbach calculation based on the D1M+QRPA+0lim  and SMLO $\gamma$SF and different nuclear level density prescriptions. All nuclear level densities \cite{Koning08,Goriely08} are normalized to the existing $s$-wave spacing data at the neutron binding energy \cite{Capote09}. As shown in Fig.~\ref{fig:ng}, the calculated cross sections are in rather good agreement with available experimental data in the keV region, although D1M+QRPA+0lim tend to underestimate the $^{136}\rm{Ba}$(n,$\gamma$)$^{137}\rm{Ba}$ cross section and SMLO to overestimate the $^{137}\rm{Ba}$(n,$\gamma$)$^{138}\rm{Ba}$ cross section.

  %---------------------------------------------------%
\begin{figure}
\includegraphics[width=1.0\columnwidth]{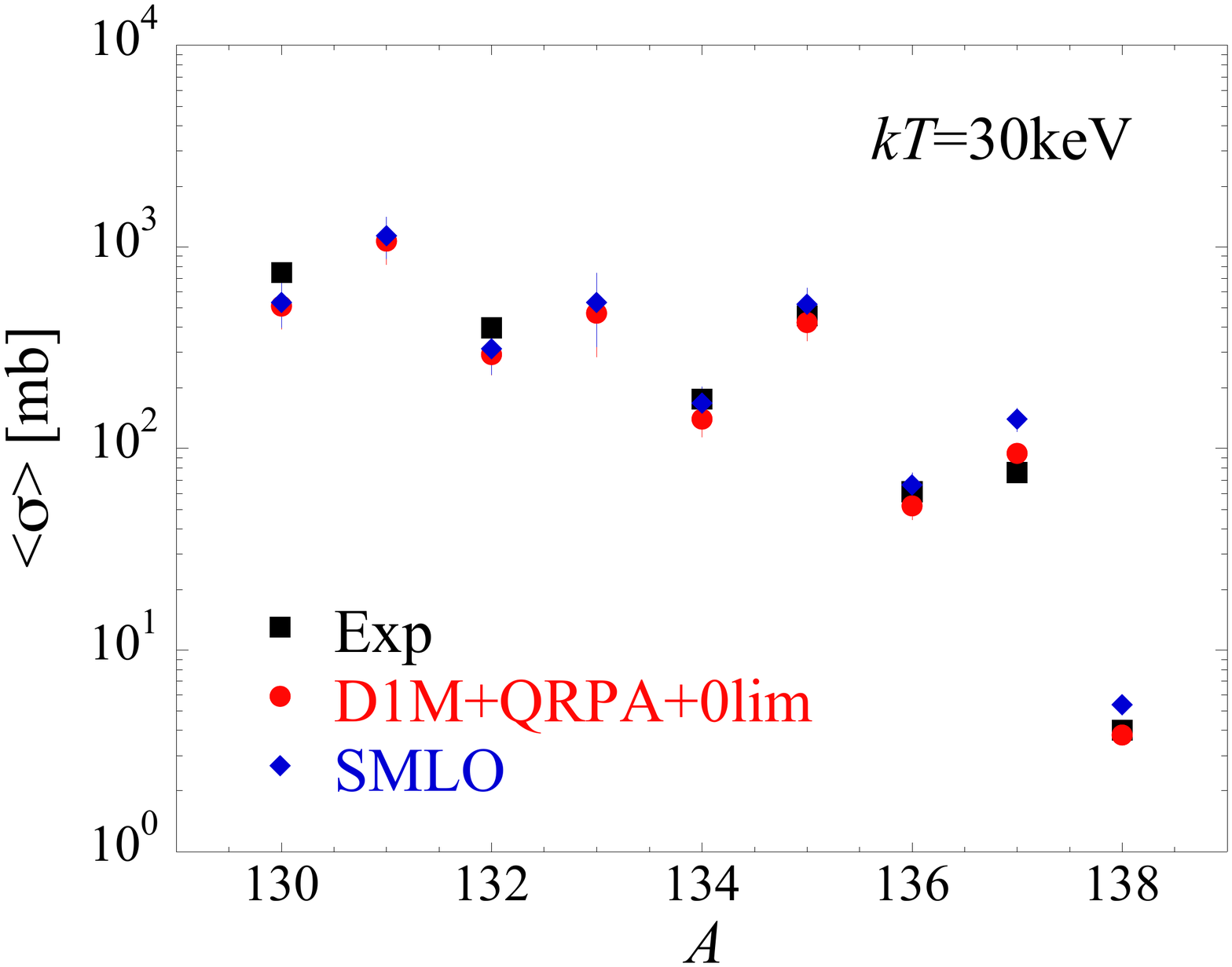}
\caption{(Color online) Comparsion between experimental (black squares) \cite{Bao00} and theoretical Maxwellian-averaged cross sections at 30keV predicted by the TALYS code with D1M+QRPA+0lim  (red circles) or SMLO (blue diamonds) $\gamma$SF for the radiative neutron capture on Ba isotopes with $A$ lying between 130 and 138.}
\label{fig:macs}
\end{figure}
%---------------------------------------------------%

To further compare the predictive power of the D1M+QRPA+0lim  and SMLO models, we compare in Fig.~\ref{fig:macs} the 30~keV experimental Maxwellian-averaged cross sections \cite{Bao00} with the TALYS predictions based again on both $\gamma$SF models and different prescriptions for the nuclear level densities \cite{Koning08,Goriely08}. For the 7 stable isotopes of Ba for which experimental values exist, the agreement is relatively good, D1M+QRPA+0lim model giving a root-mean-square deviation on the cross section ratio of 1.26 and SMLO of 1.36. The experimentally unknown Maxwellian-averaged cross sections for $^{131}$Ba and $^{133}$Ba can therefore be estimated within a typical uncertainty of 30\%. We find for $^{131}$Ba a 30~keV Maxwellian-averaged cross section of $1100 \pm 300$~mb and for $^{133}$Ba of $500 \pm 200$~mb

\section{Conclusion}
\label{sec_conc}
We presented a new experimental determination of the $(\gamma,n)$ cross section for $^{137}$Ba and $^{138}$Ba performed at the NewSUBARU synchrotron radiation facility. Our new measurements cover the low-energy tail of the giant dipole resonance above the neutron threshold and are found for the $^{138}$Ba case to be in excellent agreement with previous measurements. 
In the $^{137}$Ba case, photoneutron cross sections were measured for the first time in this experiment.  
The new cross sections were used to constrain the E1 and M1 strength functions obtained within the semi-microscopic D1M+QRPA and phenomenological SMLO approaches. We have further confirmed the relevance of the experimentally constrained D1M+QRPA dipole $\gamma$-ray strength function by analyzing the radiative neutron capture cross sections for Ba isotopes considering in addition the zero-limit systematics for both the de-excitation E1 and M1 strengths. Finally, the present analysis was used to estimate the Maxwellian-averaged  $^{131}\rm{Ba}$(n,$\gamma$)$^{132}\rm{Ba}$ and  $^{133}\rm{Ba}$(n,$\gamma$)$^{134}\rm{Ba}$ cross sections. 

%Last, but not least, the Brink hypothesis of the approximate equality between the excitation and deexcitation $\gamma$-ray strength functions is violated due to the presence of the zero-limit M1 and E1 strengths which is unique to the deexcitation $\gamma$-ray strength function.  However, a question remains; does the Brink hypothesis hold except for the zero-limit strengths? 

\section{Acknowledgments}
The authors are grateful to H. Ohgaki of the Institute of Advanced Energy, Kyoto University for making a large volume LaBr$_3$(Ce) detector available for the experiment. H.U. acknowledges the support from the Premier Project of the Konan University. S.G. acknowledges the support from the F.R.S.-FNRS. G.M.T. acknowledges funding from the Research Council of Norway, Project Grant Nos. 262952. This work was supported by the IAEA and performed within the IAEA CRP on ``Updating the Photonuclear data Library and generating a Reference Database for Photon Strength Functions'' (F41032).

\bibliographystyle{apsrev4-1}
\bibliography{newsubarubibfile}

\begin{thebibliography}{}
\bibitem{Bartholomew73} G.A. Bartholomew, E.D. Earle, A.J. Fergusson, J.W. Knowles, mad M.A. Lone, Adv. Nucl. Phys. {\bf 7}, 229 (1973).
\bibitem{Lone85} M.A. Lone, Proc. 4th Int. Symp., Smolenice, Czechoslovakia, 1985, J. Kristin, E. Betak (eds.), D. Reidel, Dordrecht, Holland (1986) 238.
\bibitem{RIPL3} R. Capote, M. Herman, P. Oblo\u zinsk\'y, P.G. Young, S. Goriely, T. Belgya, A.V. Ignatyuk, A.J. Koning, S. Hilaire, V.A. Plujko, M. Avrigeanu, O. Bersillon, M.B. Chadwick, T. Fukahori, Zhigang Ge, Yinlu Han, S. Kailas, J. Kopecky, V.M. Maslov, G. Reffo, M. Sin, E.Sh. Soukhovistskii, P. Talou, Nuclear Data Sheets {\bf 110}, 3107 (2009). 
\bibitem{Goriely18a} S. Goriely, S. Hilaire, S. P\'eru, K. Sieja, Phys. Rev. C 98 (2018) 014327.
\bibitem{Voin04}A. Voinov, E. Algin, U. Agvaanluvsan, T. Belgya, R. Chankova, M. Guttormsen, G. E. Mitchell, J. Rekstad, A. Schiller, and S. Siem, 
Phys. Rev. Lett. {\bf 93}, 142504 (2004).
\bibitem{Gutt05} M. Guttormsen, R. Chankova, U. Agvaanluvsan, E. Algin, L. A. Bernstein, F. Ingebretsen, T. L\"onnroth, S. Messelt, 
G. E. Mitchell, J. Rekstad, A. Schiller, S. Siem, A. C. Sunde, A. Voinov, and S. {\O}deg{\aa}rd, Phys. Rev. C {\bf 71}, 044307 (2005).
\bibitem{Algi08} E. Algin, U. Agvaanluvsan, M. Guttormsen, A. C. Larsen, G. E. Mitchell, J. Rekstad, A. Schiller, S. Siem, and A. Voinov,
Phys. Rev. C {\bf 78}, 054321 (2008).
\bibitem{Schw13} R. Schwengner, S. Frauendorf, and A. C. Larsen, Phys. Rev. Lett. {\bf 111}, 232504 (2013).
\bibitem{Brow14} B. A. Brown and A. C. Larsen, Phys. Rev. Lett. {\bf 113}, 252502 (2014).
\bibitem{Siej17a} K. Sieja, Phys. Rev. Lett. {\bf 119}, 052502 (2017).
\bibitem{Siej17b} K. Sieja, Europhys. J. Web Conf. {\bf 146}, 05004 (2017).
\bibitem{Kara17} S. Karampagia, B. A. Brown, and V. Zelevinsky, Phys. Rev. C {\bf 95}, 024322 (2017).
\bibitem{Schw17} R. Schwengner, S. Frauendorf, and B. A. Brown, Phys. Rev. Lett. {\bf 118}, 092502 (2017).
\bibitem{Brink} D.M. Brink, Ph.D thesis, Oxford University, 1955.
\bibitem{Axel} P. Axel, Phys. Rev. {\bf 126}, 671 (1962).
\bibitem{BW79} J.M. Blatt and V.E. Weisskopf, {\it Theoretical Nuclear Physics}, (1979 Springer-Verlag New York Inc., USA), p.761.
\bibitem{Utsu18a}H. Utsunomiya, T. Renstr{\o}m, G. M. Tveten, S. Goriely, S. Katayama, T. Ari-izumi, D. Takenaka, D. Symochko, B. V. Kheswa, V. W. Ingeberg {\it et al.}. Phy. Rev. C {\bf 98}, 054619 (2018).
\bibitem{Utsu19}H. Utsunomiya, T. Renstr{\o}m, G. M. Tveten, S. Goriely, T. Ari-izumi, D. Filipescu, J. Kaur, Y.-W. Lui, W. Luo, S. Miyamoto {\it et al.}, Phy. Rev. C {\bf 99}, 024609 (2019).
\bibitem{Goriely18b} S. Goriely, V. Plujko, Phys. Rev. C 99 (2019) 014303.
%T. Glodariu, Y.-W. Lui, S. Miyamoto, A. C. Larsen, J. E. Midtbø, A. Görgen, S. Siem, L. Crespo Campo, M. Guttormsen, S. Hilaire, S. Péru, and A. J. Koning 
\bibitem{Martini16} M. Martini, S. P\'eru, S. Hilaire, S. Goriely, and F. Lechaftois, Phys. Rev. C {\bf 94}, 014304 (2016).
\bibitem{Goriely16b} S. Goriely, S. Hilaire, S. P\'eru, M. Martini, I. Deloncle, and F. Lechaftois, Phys. Rev. C {\bf 94}, 044306 (2016).
\bibitem{TALYS} A.J. Koning, D. Rochman, Nuclear Data Sheets {\bf 113}, 2841 (2012).
%Begin experimental part
%\bibitem{MiniPIX} http://www.advacam/minipix. 
\bibitem{Utsu14} H.~Utsunomiya, T. Shima, K. Takahisa, D.M. Filipescu, O. Tesileanu, I. Gheorghe, H.-T. Nyhus, T. Renstr\o m, Y.-W. Lui, Y. Kitagawa, S. Amano, S. Miyamoto, IEEE Trans. Nucl. Sci.  {\bf 61}, 1252 (2014).
\bibitem{Ioana_thesis} A. I. Gheorghe, PhD thesis: Nuclear data obtained with Laser Compton Scattered gamma-ray beams, Ph.D. thesis, University of Bucharest (2017), unpublished.
\bibitem{Fili14} D. M. Filipescu, I. Gheorghe, H. Utsunomiya, S. Goriely, T. Renstr\o m, H.-T. Nyhus, O. Tesileanu, T. Glodariu, T. Shima, K. Takahisa, S. Miyamoto, Y.-W. Lui, S. Hilaire, S. P\'eru, M. Martini, and A. J. Koning, Phys. Rev. C {\bf 90}, 064616 (2014).
\bibitem{Utsu15} H.~Utsunomiya, S. Katayama, I. Gheorghe, S. Imai, H. Yamaguchi, D. Kahl, S. Sakaguchi, T. Shima, K. Takahisa, and S. Miyamoto, Phys. Rev. C {\bf 92}, 064323 (2015).
\bibitem{Kondo2011}T. Kondo, H. Utsunomiya, H. Akimune, T. Yama- gata, A. Okamoto, H. Harada, F. Kitatani, T. Shima, K. Horikawa, and S. Miyamoto, Nuclear Instruments and Methods in Physics Research Section A {\bf 659}, 462(2011).
\bibitem {Utsu18b} H. Utsunomiya, T. Watanabe, T. Ari-izumi, D. Takenaka, T. Araki, K. Tsuji, I. Gheorghe, D. M. Filipescu, S. Belyshev, K. Stopani, D. Symochko, H. Wang, G. Fan, T. Renstr{\o}m, G. M. Tveten, Y.-W. Lui, K. Sugita, S. Miyamoto, Nuclear Instruments and Methods in Physics Research Section A {\bf 896},  103 (2018).
\bibitem{neutrondet} O. Itoh, H. Utsunomiya, H. Akimune, T. Kondo, M. Kamata, T. Yamagata, H. Toyokawa, H. Harada, F. Kitatani, S. Goko, C. Nair, and Y.-W. Lui, Journal of Nuclear Science and Technology {\bf 48}, 834 (2011).
\bibitem{geant4ref}  J. Allison {\it et al.}, IEEE T. Nucl. Sci. {\bf 53}, 270 (2006). 
%\bibitem{Berman_ring_ratio} B. L. Berman, J. T. Caldwell, R. R. Harvey, M. A. Kelly, R. L. Bramblett, and S. C. Fultz, Phys. Rev. 162, 1098 (1967).
%\bibitem{Utsu2017} H. Utsunomiya, I. Gheorghe, D. M. Filipescu, T. Glodariu, S. Belyshev, K. Stopani, V. Varlamov, B. Ishkhanov, S. Katayama, D. Takenaka, T. Ari-izumi, S. Amano, S. Miyamoto, Nuclear Instruments and Methods in Physics Research Section A {\bf 871}, 135 (2017).
\bibitem{Rens18} T. Renstr\o m, H. Utsunomiya, H. T. Nyhus, A. C. Larsen, M. Guttormsen, G. M. Tveten, D. M. Filipescu, I. Gheorghe, S. Goriely, S. Hilaire, Y.-W. Lui, J. E. Midtb\o, S. P\'eru, T. Shima, S. Siem, and O. Tesileanu, Phy. Rev. C {\bf 98}, 054310 (2018).
%results part
\bibitem{Tonchev10} A. P. Tonchev, S. L. Hammond, J. H. Kelley, E. Kwan, H. Lenske, G. Rusev, W. Tornow, N. Tsoneva, Phy. Rev. Lett. {\bf 104}, 072501 (2010).
\bibitem{Goriely19} S. Goriely, P. Dimitriou, M. Wiedeking, {\it et. al.}, Eur. Phys. J. A (2019) submitted.
\bibitem{Berman70} B.L. Berman, S.C. Fultz, J.T. Caldwell, M.A. Kelly,  S.S. Dietrich, Phy. Rev. C {\bf 2}, 2318 (1970).
\bibitem{Varlamov16} V.V. Varlamov, B.S. Ishkhanov, V.N. Orlin, N.N. Peskov, Yadernaya Fizika {\bf 79}, 315 (2016).
\bibitem{Koning08} A.J. Koning, S. Hilaire, S. Goriely, Nucl. Phys.  A {\bf 810}, 13 (2008).
\bibitem{Goriely08} S. Goriely, S. Hilaire, and A.J. Koning, Phys. Rev. C {\bf 78}, 064307 (2008).
\bibitem{Stroud72} D.B. Stroud, D.M.H. Chan, Astrophys. J. {\bf 178}, L93 (1972).
\bibitem{Voss95} F. Voss, K. Wisshak, F. Kaeppeler, Phy. Rev. C {\bf 52}, 1102 (1995).
\bibitem{Voss94} F. Voss, K. Wisshak, K. Guber, F. Kaeppeler,  G. Reffo, Phy. Rev. C {\bf 50}, 2582 (1994).
\bibitem{Koehler98} P.E. Koehler, R.R. Spencer, K.H. Guber, R.R. Winters, S. Raman, J.A. Harvey, N.W. Hill, J.C. Blackmon, D.W. Bardayan, D.C. Larson, T.A. Lewis, D.E. Pierce, M.S. Smith, Phy. Rev. C {\bf 57}, 1558 (1998).
\bibitem{Musgrove78} A.R. DeL. Musgrove, B.J. Allen, J.W. Baldeman, R.L. Macklin, Int. Conf. on Neutr. Phys. and Nucl. Data (1978), p.449
\bibitem{Capote09} R. Capote, M. Herman, P. Oblozinsky, {\it et al.}, Nuclear Data Sheets {\bf 110}, 3107 (2009).
%\bibitem{Goriely18a} S. Goriely, S. Hilaire, S. P\'eru, K. Sieja, Phys. Rev. C 98 (2018) 014327.
%\bibitem{Koning12} A.J. Koning, D. Rochman, Nuclear Data Sheets {\bf 113}, 2841 (2012).
\bibitem{Bao00} Z.Y. Bao, H. Beer, F. K\"appeler, F. Voss, K. Wisshak, T. Rauscher, At. Data Nucl. Data Tables {\bf 75}, 1 (2000).
%\bibitem{Hilaire12} S. Hilaire, M. Girod, S. Goriely, and A.J. Koning, Phys. Rev. C {\bf 86}, 064317 (2012).

%\bibitem{Casanovas18} A. Casanovas {\it et al.}, EPJ Web of Conferences 178, 03004 (2018) (https://doi.org/10.1051/epjconf/201817803004).
%\bibitem{Plujko18} V.A. Plujko, O.M. Gorbachenko, R. Capote, P. Dimitriou, Atomic Data and Nuclear Data Tables 123 (2018) 1

\end{thebibliography}

\end{document}